\documentclass[runningheads]{llncs}
\usepackage{amsmath,amssymb,amsfonts}
\usepackage{algorithmic}
\usepackage{cite}
\usepackage{graphicx}
\usepackage{textcomp}
\usepackage{booktabs}
\usepackage{multirow}
\usepackage{makecell}
\usepackage{xcolor}
\usepackage{color, colortbl}
\usepackage{xurl,lineno,microtype}
\usepackage[hidelinks]{hyperref}

\def\BibTeX{{\rm B\kern-.05em{\sc i\kern-.025em b}\kern-.08em
    T\kern-.1667em\lower.7ex\hbox{E}\kern-.125emX}}

\definecolor{Gray}{gray}{0.9}

\begin{document}

\title{Hybrid Quantum Solvers in Production: \\ how to succeed in the NISQ era?}

\titlerunning{Hybrid Quantum Solvers in Production}

\author{Eneko Osaba\inst{1} \and
Esther Villar-Rodr\'iguez\inst{1} \and \\
Aitor Gomez-Tejedor\inst{1,2} \and
Izaskun Oregi\inst{1,3}
}

\authorrunning{Osaba et al.}

\institute{TECNALIA, Basque Research and Technology Alliance (BRTA), 48160 Derio, Spain \and
University of the Basque Country UPV/EHU, 48940 Leioa, Spain \and
European University of Gasteiz, EUNEIZ, 01013 Vitoria-Gasteiz, Spain\\
\email{eneko.osaba@tecnalia.com}}

\maketitle

\begin{abstract}
Hybrid quantum computing is considered the present and the future within the field of quantum computing. Far from being a passing fad, this trend cannot be considered just a stopgap to address the limitations of NISQ-era devices. The foundations linking both computing paradigms will remain robust over time. The contribution of this work is twofold: first, we describe and categorize some of the most frequently used hybrid solvers, resorting to two different taxonomies recently published in the literature. Secondly, we put a special focus on two solvers that are currently deployed in real production and that have demonstrated to be near the real industry. These solvers are the \texttt{LeapHybridBQMSampler} contained in D-Wave's Hybrid Solver Service and Quantagonia's Hybrid Solver. We analyze the performance of both methods using as benchmarks four combinatorial optimization problems.

\end{abstract}

\section{Introduction}
Recent years have seen significant progress in quantum computing (QC), primarily as a result of the fast development of the technology and the advances made in the democratization of its access \cite{seskir2023democratization}. As a result, QC has already helped several application domains launch various proofs of concept. Frequently, these endeavors depend on hybrid systems, which symbolize the immediate future of this field. This is so since hardware capabilities play a major role in the adoption of quantum approaches to solve industrial use cases. Being immersed in the \textit{noisy intermediate scale quantum} (NISQ, \cite{preskill2018quantum}) era, today's quantum technology is neither entirely dependable nor competent to resolve medium- or large-sized problems. In this context, the challenge lies in finding the way to build fruitful harmony between classical and quantum computing, assembling a tandem that, eventually, outperforms previous purely classical conceptualizations.

Furthermore, it would be a mistake to conceive of quantum-classical hybrid computing simply as a stopgap to address the limitations of NISQ-era computers, just as quantum computing should not be referred to as a natural substitute for classical computing. As stated in \cite{callison2022hybrid}, ``\textit{hybrid algorithms are likely here to stay well past the NISQ era and even into full fault-tolerance, with the quantum processors augmenting the already powerful classical processors which exist by performing specialized task}". In other words, classical and quantum computing will play collaborative and cooperative (but never competitive) roles in the mid-term horizon.

\begin{figure}[t]
	\centering
	\includegraphics[width=0.6\linewidth]{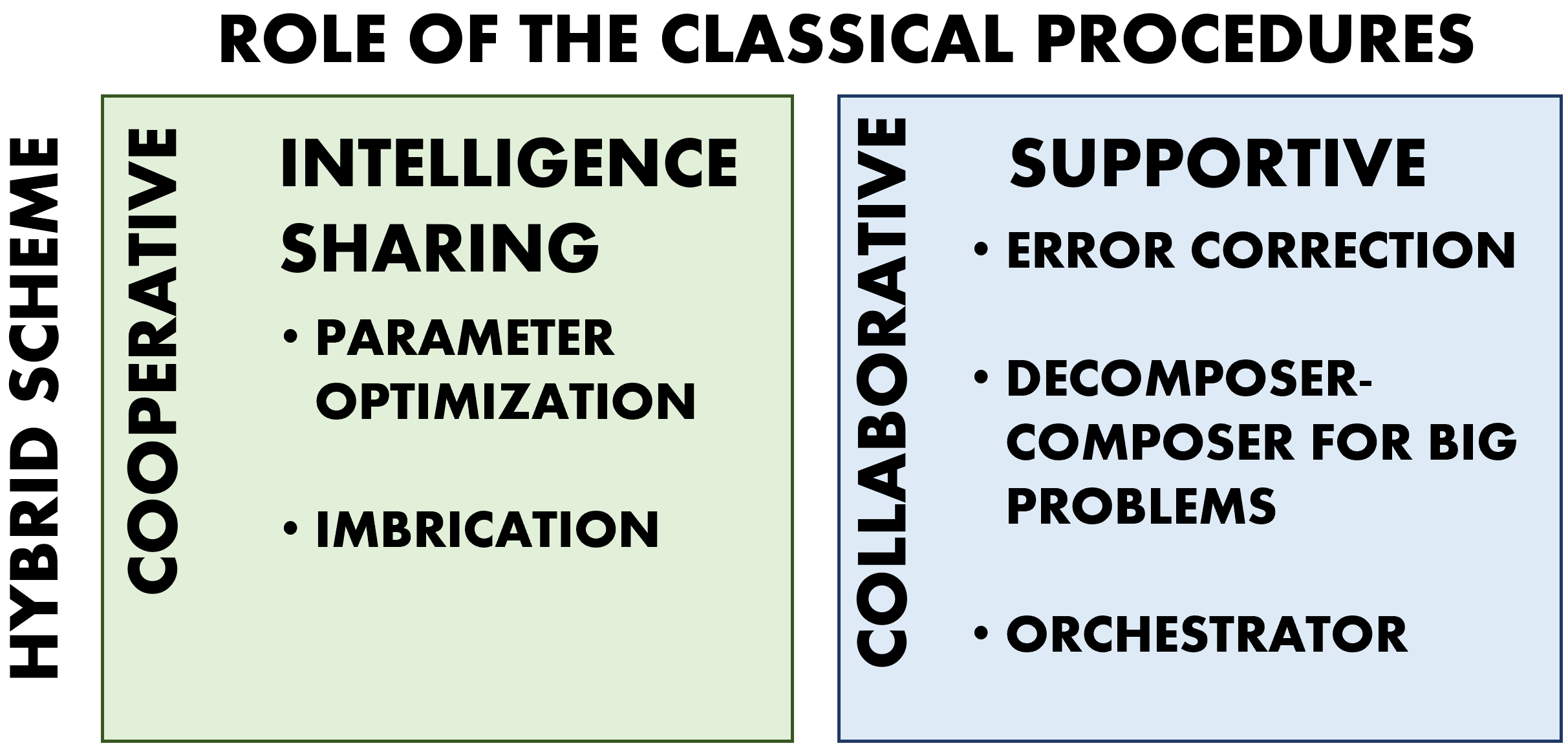}
	\caption{Classification of HS schemes. Image based on \cite{villar2023hybrid}.}
	\label{fig:tlp}
\end{figure}

Despite the work carried out in this area, there are a multitude of aspects that deserve much more in-depth study than has been done to date. These open issues range from the proper characterization of the hybrid quantum computing world to the concrete definition of what a hybrid solver is. Specifically, in this research, main contributions revolve around:

\begin{itemize}
	\item Describing and categorizing some of the most frequently used hybrid solvers to provide a clear picture of the dominant trends in today's scientific and industrial community:
	\begin{enumerate}
		\item According to the \textit{layout established by the workflow of the hybrid solver, \textbf{vertical} versus \textbf{horizontal}}, following the guidelines of \cite{phillipson2023classification}. This means focusing on the logical arrangement of the quantum and classical modules within the method. Note that this work focuses on giving an overview of the hybrid solver strategies with emphasis on the \textit{designable-by-designer} steps. This means that the taxonomy will not include any consideration with respect to vertical hybrid workflows, i.e. those with ``\textit{controlling activities required to control and operate}" either a classical or a quantum subroutine, as these operations are not key functional stages of the algorithmic scheme of the solver.  
		
		\item According to the \textit{role of the classical mechanisms in the solving procedure: \textbf{collaborative} versus \textbf{cooperative}}, as expressed by Figure \ref{fig:tlp} brought from the work of \cite{villar2023hybrid}. This entails paying attention to the nature or motivation for the conjunction of both quantum and classical paradigms in the solver pipeline: \textit{are both in charge of finding the solution? are they sharing intelligence?} Note that, in this case, many levels of abstraction could be applied, especially with architectures relying on parallel computing. For the sake of brevity, this work reviews just the strategy of the solver as a whole without delving into the nature of each independent branch.
		
		\item Last but not least, this work provides a more detailed classification according to the contribution of their modules (or blocks of modules) regarding \textit{the main functional stages of an optimization algorithm, i.e. \textbf{exploration} and \textbf{exploitation} phases}. The objective is to pinpoint the processes where the community believes the quantum method can provide the most value.
	\end{enumerate} 
	
	\item Given the previous theoretical framework, gaining insight into the performance of those hybridization strategies nowadays in production and close to industrial markets. To do that, an experimentation on two hybrid solvers is conducted: the \texttt{LeapHybridBQMSampler} (\texttt{LeapBQM}) contained in D-Wave's Hybrid Solver Service (HSS, \cite{HSS}), and Quantagonia's Hybrid Solver\footnote{\url{https://www.quantagonia.com/hybridsolver}}. For these tests, the QOPTLib benchmark \cite{osaba2023qoptlib} has been employed, which is composed of 40 instances equally distributed over four combinatorial optimization problems: the Traveling Salesman Problem (TSP), the Vehicle Routing Problem (VRP), the one-dimensional Bin Packing Problem (BPP), and the Maximum Cut Problem (MCP).
\end{itemize}

The rest of this paper is structured as follows: In the following Section~\ref{sec:where}, we perform the classification of five of the most renowned and used hybrid solvers to summarize current trends for hybrid solver designs, which answers the main question posed in the title of the section: ``\emph{Where might the quantum advantage be?}". After that, Section~\ref{sec:exp} describes the experimentation conducted with \texttt{LeapBQM} and \texttt{QHS}, including the experimental setup and results. Finally, Section \ref{sec:conc} closes this document by drawing some conclusions and outlining future research lines.

\section{Classifying hybrid solvers: Where might the quantum advantage be?} \label{sec:where}

A myriad of hybrid solvers have been proposed in the literature in recent years. Many of these methods are developed ad-hoc for solving a specific use case, as can be seen in examples such as \cite{ajagekar2022hybrid,osaba2021hybrid,gao2023solving}. Typically, researchers design their hybrid solvers taking into account several aspects, as \textit{i)} the characteristics and limitations of the available quantum resources; \textit{ii)} the particularities and constraints of the problem to be addressed; or \textit{iii)} the intuition and knowledge of the researcher; among many others.

However, there are a number of perfectly recognizable schemes that have been frequently employed by the community to deal with optimization problems. From the perspective of the work in \cite{phillipson2023classification}, all of these configurations are \textbf{horizontal hybrid} schemes since they are pipelines containing ``\textit{all operational activities required to use a quantum computer and a classical computer to perform an algorithm}".

The following are examples of well-known \textbf{micro hybrid split} schemes, that is, solvers that deal with a single activity, in which some operations are quantum and others are classical (in an interative fashion):

\begin{itemize}
	\item \textbf{Variational Quantum Algorithms (VQAs)}: VQAs are the most used hybrid schemes for solving optimization problems through the gate-based quantum paradigm. As briefly explained in \cite{cerezo2021variational}, ``\textit{the trademark of VQAs is that they use a quantum computer to estimate the cost function of a problem (or its gradient) while leveraging the power of classical optimizers to train the parameters of the quantum circuit}". The most representative examples of the VQA framework are the Variational Quantum Eigensolver (VQE, \cite{peruzzo2014variational}) and the Quantum Approximate Optimization Algorithm (QAOA, \cite{farhi2014quantum}). We refer readers interested in VQE and QAOA applications to \cite{tilly2022variational} and \cite{blekos2023review}. 
	
	Following the taxonomy presented in \cite{villar2023hybrid} and represented in Figure \ref{fig:tlp}, VQAs can be considered as hybrid schemes with a classical method optimizing the circuit parameters, thus playing a \textbf{cooperative} role where the solver ``\textit{contains a core intelligence engine composed of both classical and quantum artifacts}”. Furthermore, in VQAs, there is no clear division of exploration and exploitation activities.
	
	Finally, Figure \ref{fig:vqa} describes the general workflow of a VQA, specifying which step is carried out in a quantum device and which in a classical computer.
	
	\begin{figure}[t]
		\centering
		\includegraphics[width=0.8\linewidth]{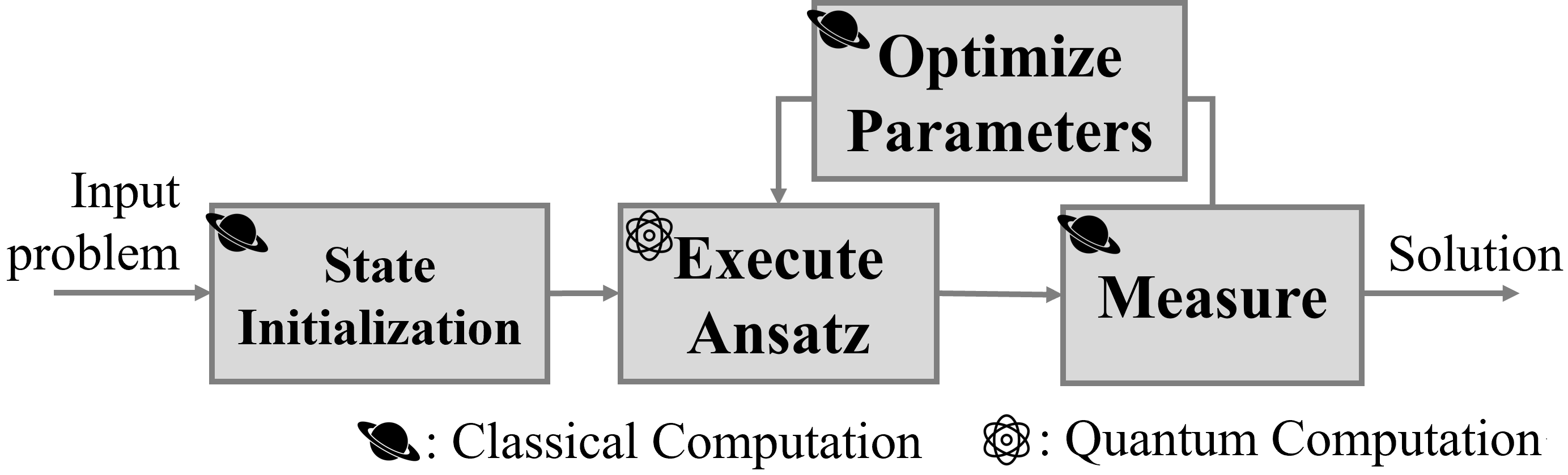}
		\caption{General workflow of VQA solvers.}
		\label{fig:vqa}
	\end{figure}
	
	\item \textbf{QBSolv}: QBSolv is a hybrid method developed by D-Wave, which can be described as ``\textit{a decomposing solver for solving a QUBO problem by splitting it into pieces}"\footnote{\url{https://github.com/dwavesystems/qbsolv}}. The decomposing procedure of this method starts with an initial solution provided by a classical Tabu Search algorithm \cite{glover1990tabu}. A quantum method later improves that solution by sending a user-specified \textit{fraction} of it to the QPU. This subproblem includes the \textit{fraction} of most energetic variables of the initial problem. In practice, this value is usually between 0.05 and 0.15.
	
	Despite the fact this technique was deprecated as of the end of 2021 and discontinued after March 2022, QBSolv has been widely used by researchers, even in recent studies \cite{teplukhin2022sampling,wang2022research,tosun2022new}. Embracing the Figure \ref{fig:tlp} taxonomy, the quantum and classical procedures pull together with an imbricated, hence \textbf{cooperative}, scheme where ``\textit{both participate in searching for the solution of the problem}" \cite{villar2023hybrid}. Figure \ref{fig:qbsolv} displays the general workflow of QBSolv.
	
	\begin{figure}[t]
		\centering
		\includegraphics[width=0.55\linewidth]{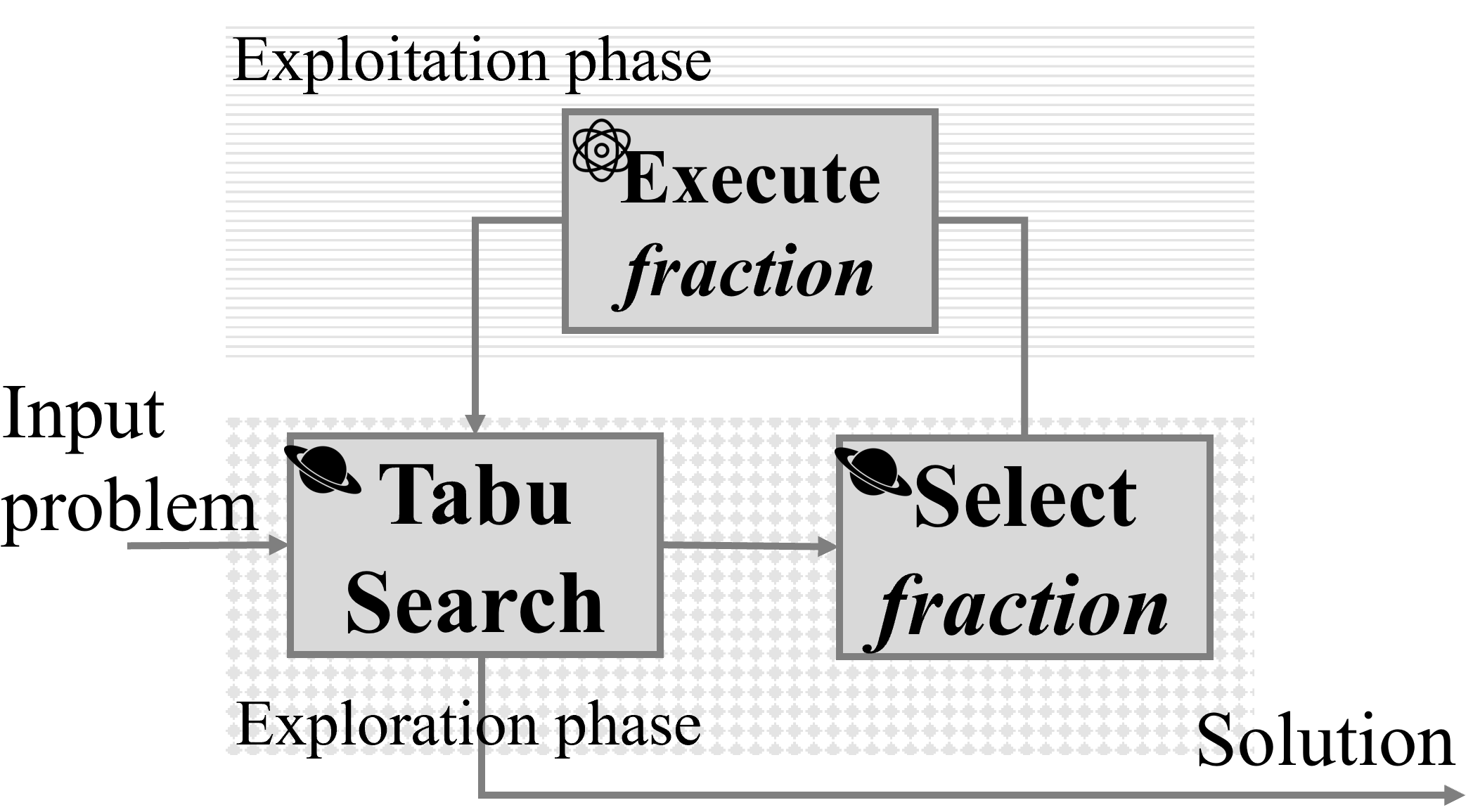}
		\caption{General workflow of QBSolv.}
		\label{fig:qbsolv}
	\end{figure}
	
\end{itemize}

One step forward from these standard hybrid schemes is taken when introducing parallel computing in advanced hybrid solver designs. Accordingly and in conformity with the criteria set out in \cite{phillipson2023classification}, the following solvers fall within the \textbf{parallel hybrid} category since ``\textit{the workflows of parallel hybrid have multiple independent branches to solve a specific problem}". Arguably, the most representative example is the  \textbf{D-Wave-Hybrid-Framework \cite{Hybrid}}.

\begin{figure}[t]
	\centering
	\includegraphics[width=0.9\linewidth]{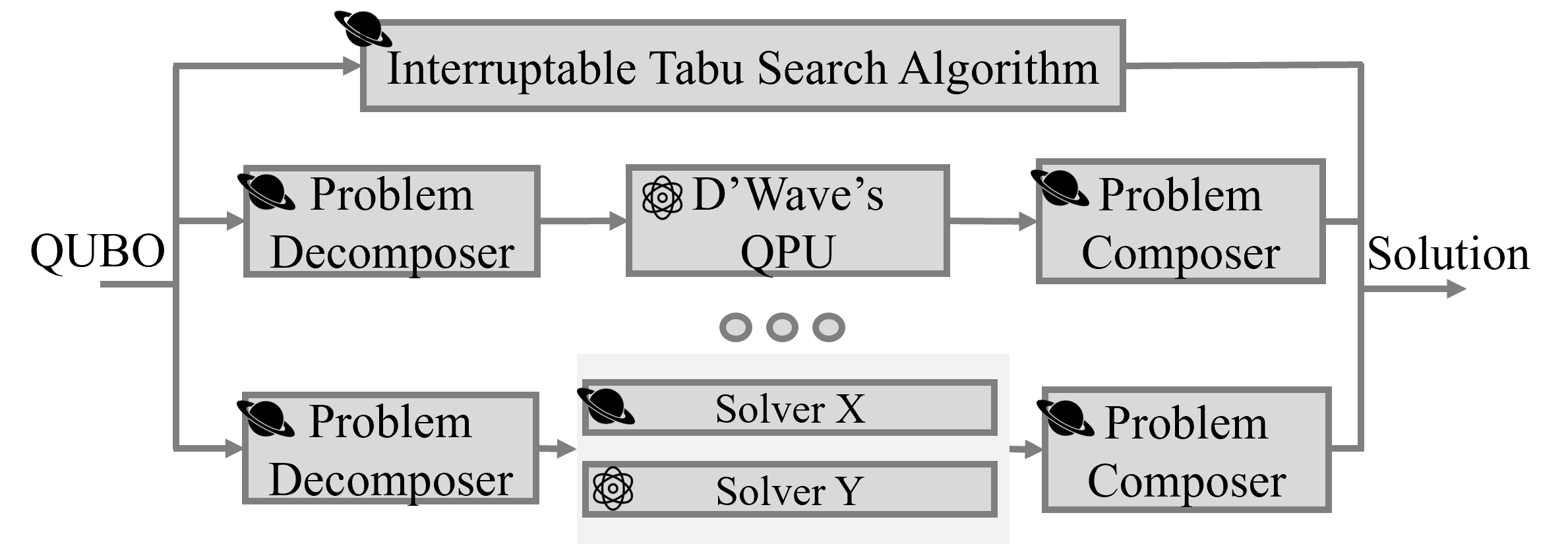}
	\caption{General scheme of D-Wave-Hybrid-Framework.}
	\label{fig:Hybrid}
\end{figure}

As explained by its creators, the D-Wave-Hybrid-Framework is ``\textit{a general, minimal Python framework for building hybrid asynchronous decomposition samplers for QUBO problems}"\footnote{\url{https://docs.ocean.dwavesys.com/en/stable/docs_hybrid/}}. This framework is appropriate for ``\textit{developing hybrid approaches to combining quantum and classical compute resources}". Figure \ref{fig:Hybrid} represents the general scheme of the D-Wave-Hybrid-Framework as it was conceived in \cite{Hybrid}. The best known concretization is the solver coined as \textit{Kerberos}. This solver is a reference hybrid workflow which is composed of three different techniques running in parallel for a number of iterations: a quantum one that accesses the QPU and two classical ones, a Tabu and a Simulated Annealing \cite{bertsimas1993simulated}. It should be noted that these branches share solutions along the execution. That is, the best solution found by the branches in an iteration is fed into all the branches in the next iteration. Recent practical applications of \textit{Kerberos} can be found in papers such as \cite{malviya2023logistics,stogiannos2022experimental}. 

Regarding the taxonomy in Figure \ref{fig:tlp}, and because the best solution found by all branches is shared, and thereby improved, among the counterparts, D-Wave-Hybrid-Framework can be conceived as an imbricated solver with a \textbf{cooperative} strategy.

Lately, some concrete efforts have been put forth to bring hybrid solvers to industry. This means giving the users the opportunity of releasing hybrid solvers more straightforwardly while assuring a good performance. In this article, we focus on describing and analyzing two approaches of this kind, which have demonstrated to be close to the real market: D-Wave's Hybrid Solver Service and Quantagonia's Hybrid Solver.

\begin{figure}[t]
	\centering
	\includegraphics[width=0.5\linewidth]{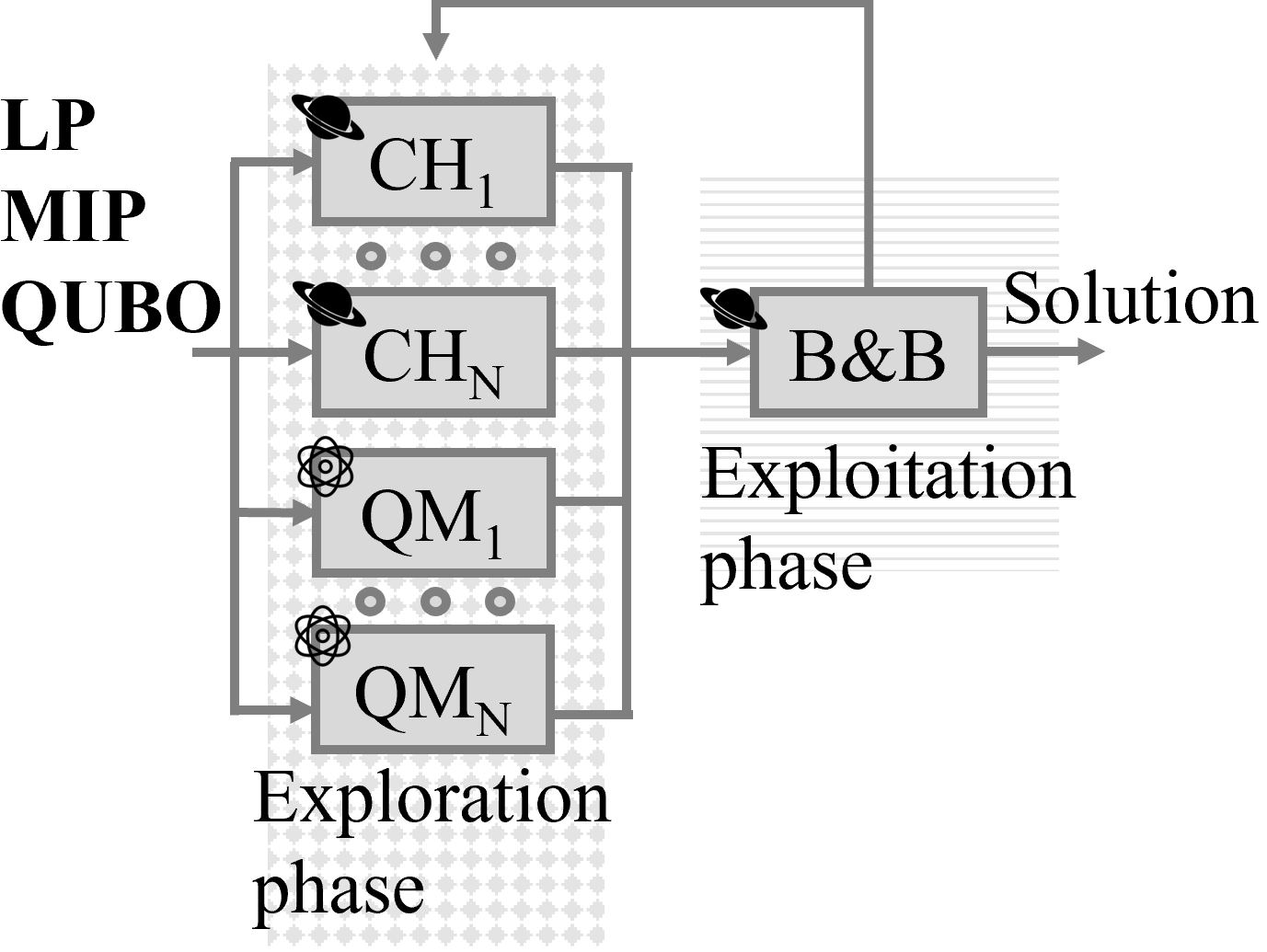}
	\caption{Quantagonia's Hybrid Solver workflow. CH = classical heuristic. QM = Quantum Module. B\&B = Branch and Bound algorithm.}
	\label{fig:quantagonia}
\end{figure}

\begin{itemize}
	\item \textbf{D-Wave's Hybrid Solver Service}: HSS consists of a portfolio of hybrid heuristic solvers that leverage classical and quantum computation to address large optimization problems \cite{glos2023optimizing,colucci2023power,v2023hybrid}, or for real-world industrial cases\footnote{\url{https://www.dwavesys.com/learn/resource-library/}}. At the time of this writing, the HSS accommodates three different solvers for dealing with three problem types \cite{leapCQM}: the binary quadratic model (BQM) solver, \texttt{LeapBQM}, for problems defined on binary values; the discrete quadratic model (DQM) method, \texttt{LeapDQM}, for problems defined on discrete values; and the constrained quadratic model (CQMs) technique, \texttt{LeapCQM}, which can face problems defined on binary, integer, and even real values.
	
	Each solver in HSS has the same structure, which is depicted in Figure \ref{fig:hss}:
	
	\begin{enumerate}
		\item First, the method starts reading the input problem, accepting BQM, DQM or CQM formats. 
		\item Next, HSS generates one or more hybrid threads devoted to solving the problem at hand. Each thread counts with a Classical Heuristic Module (CH), which is in charge of exploring the complete solution space, and a Quantum Module (QM), which, in the words of D-Wave, ``\textit{formulates quantum queries that are sent to a back-end Advantage QPU. Replies from the QPU are used to guide the heuristic module toward more promising areas of the search space or to find improvements to existing solutions}". In other words, QM is mainly in charge of exploitation duties. Finally, the solver forwards to the user the best solution among those found in the pool of threads.
	\end{enumerate}
	
	\begin{figure}[t]
		\centering
		\includegraphics[width=0.45\linewidth]{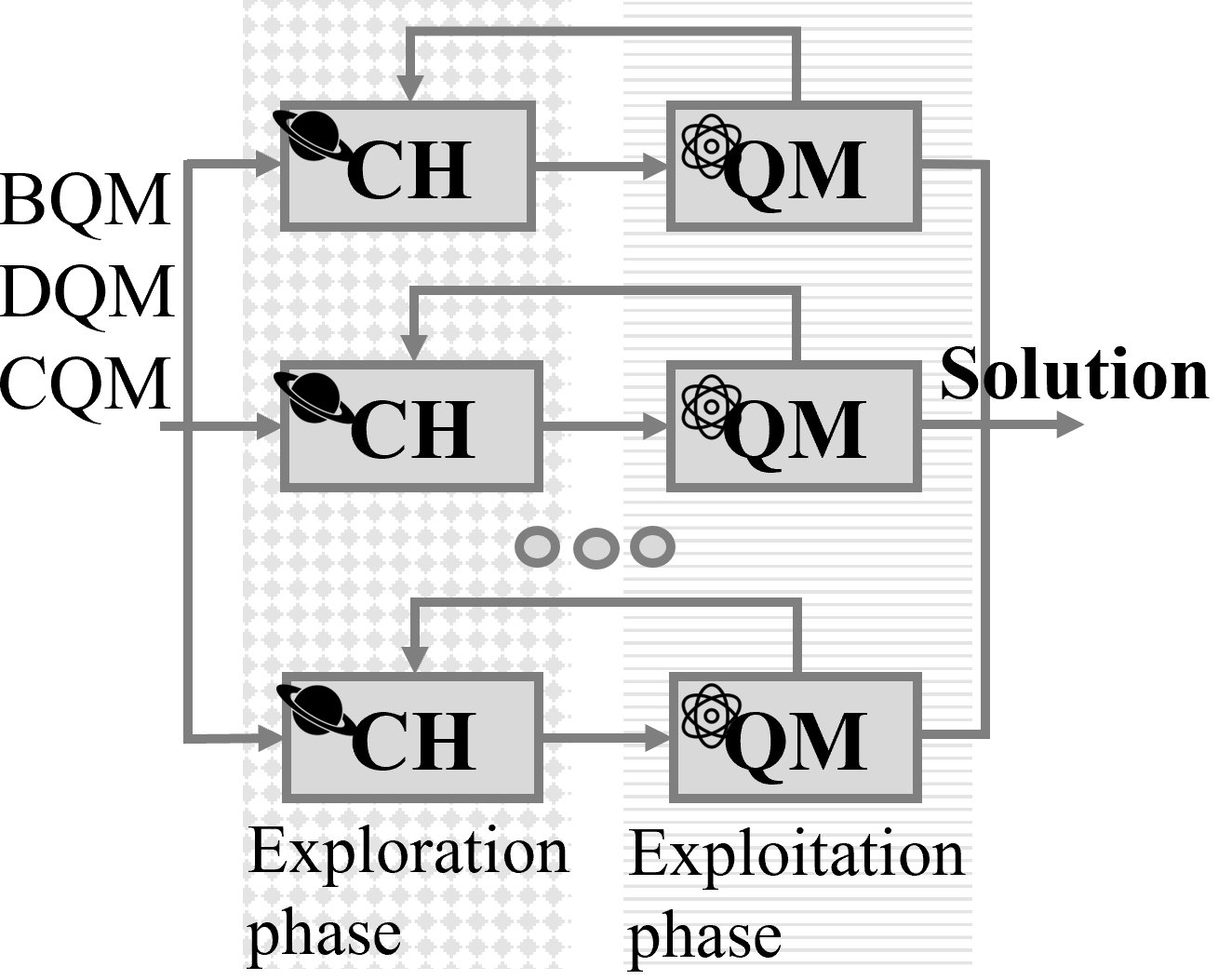}
		\caption{D-Wave's Hybrid Solver Service schemes. CH = Classical Heuristic Module. QM = Quantum Module.}
		\label{fig:hss}
	\end{figure}
	
	Given this definition, HSS belongs to the \textbf{cooperative} hybrid category, and more concretely to the imbricated schemes.
	
	\item \textbf{Quantagonia's Hybrid Solver}: established in November 2021, Quantagonia's main objective is to bridge the gap between quantum computing, mainly in labs, and real-world industry needs. Among the portfolio of products offered by this German company, \texttt{QHS} is a hardware-agnostic method for addressing Linear Programming, Mixed-Integer Programming and QUBO problems. Despite the fact that \texttt{QHS} has not been employed in any research paper yet, it is currently being used by leading companies such as Strangeworks, MathPlan, T-Systems, and Adesso, which reinforces its practical application.
	
	Regarding the solver's anatomy, it is divided into two distinguishable steps, as shown in Figure \ref{fig:quantagonia}:
	
	\begin{itemize}
		\item First, \texttt{QHS} runs a set of \textit{primal heuristics} in parallel that solve the complete problem. These primal heuristics can be either classical, such as a Simulated Annealing, or quantum by accessing external services like the D-Wave's \textit{Advantage\_System} QPU. 
		
		\item Then, \texttt{QHS} improves the best solution found by the primal heuristics by means of a classical Branch-and-Bound algorithm \cite{lawler1966branch}. It is interesting to mention that primal heuristics are still applied while the Branch and Bound method is running in an attempt to improve the best solution found so far.
		
	\end{itemize}
	
	Furthermore, and being one of its strengths, \texttt{QHS} includes an \textit{optimality proof} mechanism, which means that, along with the results, the solver also provides the \textit{optimality gap}. The \textit{optimality gap} tells how much more potential there could be in the optimization process. This gap would be 0\% when the solution found is proven to be the optimal one. Another advantage of \texttt{QHS} in contrast with HSS lies in its flexibility and ample configurability. Despite being a solver conceived to be in production, \texttt{QHS} gives a greater control to the user, such as the (de)activation of the Branch-and-Bound step, the addition of further primal heuristics, or the setting of a time limit, among many others. On the contrary, HSS methods are blackbox algorithms just allowing a maximum runtime to be set.
	
	Following the taxonomy proposed in \cite{villar2023hybrid}, \texttt{QHS} can also be conceived as an imbricated method with a classical mechanism \textbf{cooperating} to find a solution, as represented in Figure \ref{fig:quantagonia}. 
	
	It is worth pointing out the difference between \texttt{QHS} and HSS with respect to the role played by the quantum module. While the exploitation stage is mainly governed by quantum computing in the majority of hybrid solvers in the literature, \texttt{QHS} exchanges the roles proposing the quantum module as the routine responsible for the exploration.
	
\end{itemize}

Lastly, we depict in Table \ref{tab:cat} a summary of the complete categorization described along this section.

\begin{table*}[!t]
	\centering
	\caption{Categorization of the hybrid solvers described}\label{tab:exp}
	\resizebox{1.0\columnwidth}{!}{
		\begin{tabular}{cccc}
			\toprule[1.5pt]
			
			\textbf{SOLVER} & \textbf{CLASSIFICATION BY \cite{villar2023hybrid}} 
			& \makecell{\textbf{Responsible for EXPLORATION} \\ \textbf{and EXPLOITATION}}\\
			\toprule[1.5pt]
			
			\textbf{VQA} & \textbf{Cooperative - Parameter Optimization} 
			& Responsibility shared \\ \midrule
			
			\textbf{QBSolv} & \textbf{Cooperative - Imbrication} 
			& \makecell{Exploration: classical computing, \\  Exploitation: QPU}  \\ \midrule

			\makecell{\textbf{D-Wave-Hybrid}- \\ \textbf{Framework}} & \textbf{Cooperative - Imbrication} 
			& Responsibility shared \\ \midrule
			
			\textbf{HSS} & \makecell{\textbf{Cooperative - Imbrication} \\ (each generated thread)} 
			& \makecell{Exploration: classical computing, \\  Exploitation: QPU} \\ \midrule
			
			\textbf{QHS} & \textbf{Cooperative - Imbrication} 
			& \makecell{Exploration: QPU, \\ Exploitation: classical computing} \\
			
			\bottomrule[1.25pt]
		\end{tabular}
	}
	\label{tab:cat}
\end{table*}

\section{Experimentation} \label{sec:exp}

In order to evaluate the performance in terms of quality of both \texttt{LeapBQM} and \texttt{QHS}, a comprehensive experimental setup has been designed over a combinatorial optimization benchmark coined QOPTLib \cite{osaba2023qoptlib}. Two are the main reasons that have encouraged us to employ \texttt{LeapBQM} and \texttt{QHS} for the experimentation: 

\begin{enumerate}
	\item \texttt{LeapBQM} and \texttt{QHS} have successfully demonstrated to be near the applied industry, being currently used by some leading companies of different types. Delving into advanced solvers such as these provides the reader with an approximate idea of the performance and philosophy of methods that are currently deployed in real production.
	
	\item Both solvers are built under different design principles, meaning that the QPUs are used for different purposes. This fact allows the reader to analyze different implementation pathways and the possible impact that they may have on the overall performance of the algorithm.
\end{enumerate}

As mentioned, the complete QOPTLib has been employed for benchmarking purposes. In a nutshell, QOPTLib is a quantum computing-oriented benchmark for combinatorial optimization problems, and it is comprised of 40 instances equally divided into four well-known problems: 

\begin{enumerate}
	\item \textbf{TSP}, for which each instance is named as \texttt{wiX} or \texttt{djX}, where \texttt{X} represents the number of nodes to visit. For the TSP, the objective function represents the total distance of the calculated route.
	\item \textbf{VRP}, for which each case is coined \texttt{P-nX\_Y}, where \texttt{X} is the number of clients to attend and \texttt{Y} is a suffix to distinguish the set of instances with the same \texttt{X}. As for the TSP, the objective function of the VRP is measured as the total distances of the routes that compose a solution.
	\item \textbf{BPP}, with instances called \texttt{BPP\_X}, with \texttt{X} depicting the number of packages to store. The objective function for the BPP is the number of bins required to store the items that make up the instance.
	\item \textbf{MCP}, for which each case is named \texttt{MaxCut\_X}, being \texttt{X} the number of nodes that define the graph. The objective function of the MCP is calculated by aggregating the weights of the cut edges.
\end{enumerate}

Regarding the parameterization, the default values have been used for both solvers. Furthermore, for \texttt{QHS}, in addition to the default primal heuristics, the \textit{D-Wave Simulator} heuristic has also been included for running these quantum-oriented tests. As for the method versions, v2.2 of \texttt{LeapBQM} and v1.1.1145 of \texttt{QHS} have been employed. Finally, for building the corresponding QUBOs, Qiskit v0.6.0 open libraries have been resorted to for the four optimization problems\footnote{\url{https://qiskit.org/ecosystem/optimization/apidocs/qiskit_optimization.applications.html}}.

Thus, Table \ref{tab:results} shows the results obtained after 10 independent runs per instance. For each instance, we represent the result average obtained for each solver, as well as the standard deviation and the median value. Note that all these values correspond to the original objective functions above described. Aiming to enhance the replicability of this work, all the 40 instances used and the results obtained are openly available in \cite{PaperRep}.

\begin{table*}[t!]
	\centering
	\caption{Average, Standard Deviation and Median of the results obtained by \texttt{LeapBQM} and \texttt{QHS} for the whole \texttt{QOptLib} benchmark. Note that the values correspond to the original objective function already especified in Section \ref{sec:exp}. Best average values per instance have been highlighted in \textbf{bold}.}
	\resizebox{1.0\columnwidth}{!}{
		\begin{tabular}{c|lll|lll|c|lll|lll}
			\toprule[1.5pt]
			
			& \multicolumn{6}{c}{\bf Traveling Salesman Problem} & \multicolumn{7}{c}{ \bf Bin Packing Problem}\\\midrule
			\multirow{2}{*}{\bf Instance} & \multicolumn{3}{c}{\texttt{LeapBQM}} & \multicolumn{3}{c}{\texttt{QHS}} & \multirow{2}{*}{\bf Instance} & \multicolumn{3}{c}{\texttt{LeapBQM}} & \multicolumn{3}{c}{\texttt{QHS}}\\
			
			& Av. & St. & Median & Av. & St. & Median & & Av. & St. & Median & Av. & St. & Median\\
			
			\midrule
			\texttt{wi4} & \textbf{6700.0}  & 0.0 & 6700.0 & \textbf{6700.0}  & 0.0 & 6700.0 & \texttt{BPP\_3} & \textbf{2.0} & 0.0 & 2.0 & \textbf{2.0} & 0.0 & 2.0\\
			\texttt{wi5} & \textbf{6786.0}  & 0.0 & 6786.0 & \textbf{6786.0}  & 0.0 & 6786.0 & \texttt{BPP\_4} & \textbf{2.0} & 0.0 & 2.0 & \textbf{2.0} & 0.0 & 2.0\\
			\texttt{wi6} & \textbf{9815.0} & 0.0 & 9815.0 & \textbf{9815.0} & 0.0 & 9815.0 & \texttt{BPP\_5} & 2.2 & 0.40 & 2.0 & \textbf{2.0} & 0.0 & 2.0\\
			\texttt{wi7} & \textbf{7245.0} & 0.0 & 7245.0 & \textbf{7245.0} & 0.0 & 7245.0 &  \texttt{BPP\_6} & 3.3 & 0.46 & 2.0 & \textbf{3.0} & 0.0 & 3.0\\
			\texttt{dj8} & 2804.1 & 55.97 & 2787.0 & \textbf{2762.0} & 0.0 & 2762.0 & \texttt{BPP\_7} & 3.7 & 0.46 & 4.0 & \textbf{3.0} & 0.0 & 3.0\\
			\texttt{dj9} & 2317.7 & 60.91 & 2309.5 & \textbf{2134.0} & 0.0 & 2134.0 & \texttt{BPP\_8} & 4.1 & 0.54 & 4.0 & \textbf{2.0} & 0.0 & 2.0\\
			\texttt{dj10} & 3016.6 & 100.24 & 3019.5 & \textbf{2822.0} & 0.0 & 2822.0 & \texttt{BPP\_9} & 4.7 & 0.90 & 5.0 & \textbf{3.0} & 0.0 & 3.0\\
			\texttt{dj15} & 6222.5 & 591.59 & 6450.5 & \textbf{3237.0} & 0.0 & 3237.0 & \texttt{BPP\_10} & 6.4 & 0.80 & 6.0 & \textbf{4.0} & 0.0 & 4.0\\
			\texttt{dj22} & 11139.7 & 805.77 & 11234.5 & \textbf{4105.0} & 0.0 & 4105.0 & \texttt{BPP\_12} & 7.1 & 0.83 & 7.0 & \textbf{5.0} & 0.0 & 5.0\\
			\texttt{wi25} & 95081.3 & 10633.78 & 96942.0 & \textbf{26444.0} & 0.0 & 26444.0 & \texttt{BPP\_14} & 8.0 & 0.63 & 8.0 & \textbf{4.0} & 0.0 & 4.0\\
			
			\midrule
			& \multicolumn{6}{c|}{\bf Vehicle Routing Problem} & \multicolumn{7}{c}{\bf Maximum Cut Problem}\\\midrule
			\multirow{2}{*}{\bf Instance} & \multicolumn{3}{c|}{\texttt{LeapBQM}} & \multicolumn{3}{c|}{\texttt{QHS}} & \multirow{2}{*}{\bf Instance} & \multicolumn{3}{c|}{\texttt{LeapBQM}} & \multicolumn{3}{c}{\texttt{QHS}}\\
			
			& Av. & St. & Median & Av. & St. & Median & & Av. & St. & Median & Av. & St. & Median\\
			
			\midrule
			\texttt{P-n4\_1} & \textbf{97.0} & 0.0 & 97.0 & \textbf{97.0} & 0.0 & 97.0 & \texttt{MaxCut\_10} & \textbf{25.0} & 0.0 & 25.0 & \textbf{25.0} & 0.0 & 25.0\\
			\texttt{P-n4\_2} & \textbf{121.0} & 0.0 & 121.0 & \textbf{121.0} & 0.0 & 121.0 & \texttt{MaxCut\_20} & \textbf{97.0} & 0.0 & 97.0 & \textbf{97.0} & 0.0 & 97.0 \\
			\texttt{P-n5\_1} & \textbf{94.0} & 0.0 & 94.0 & \textbf{94.0} & 0.0 & 94.0 & \texttt{MaxCut\_40} & \textbf{355.0} & 0.0 & 355.0 & \textbf{355.0} & 0.0 & 355.0\\
			\texttt{P-n5\_2} & \textbf{295.0} & 0.0 & 295.0 & \textbf{295.0} & 0.0 & 295.0 & \texttt{MaxCut\_50} & \textbf{602.0} & 0.0 & 602.0 & \textbf{602.0} & 0.0 & 602.0\\
			\texttt{P-n6\_1} & \textbf{118.0} & 0.0 & 118.0 & \textbf{118.0} & 0.0 & 118.0 & \texttt{MaxCut\_60} & \textbf{852.0} & 0.0 & 852.0 & \textbf{852.0} & 0.0 & 852.0\\
			\texttt{P-n6\_2} & \textbf{122.0} & 0.0 & 122.0 & \textbf{122.0} & 0.0 & 122.0 & \texttt{MaxCut\_100} & \textbf{2224.0} & 0.0 & 2224.0 & \textbf{2224.0} & 0.0 & 2224.0\\
			\texttt{P-n7\_1} & 129.2 & 6.35 & 132.5 & \textbf{118.0} & 0.0 & 118.0 & \texttt{MaxCut\_150} & \textbf{4899.0} & 0.0 & 4899.0 & \textbf{4899.0} & 0.0 & 4899.0\\
			\texttt{P-n7\_2} & 147.8 & 6.85 & 147.0 & \textbf{137.6} & 1.2 & 137.0 & \texttt{MaxCut\_200} & \textbf{8717.0} & 0.0 & 8717.0 & \textbf{8717.0} & 0.0 & 8717.0\\
			\texttt{P-n8\_1} & 148.3 & 3.07 & 148.0 & \textbf{140.5} & 3.96 & 140.0 & \texttt{MaxCut\_250} & \textbf{13460.0} & 0.0 & 13460.0 & \textbf{13460.0} & 0.0 & 13460.0\\
			\texttt{P-n8\_2} & 250.0 & 13.02 & 251.0 & \textbf{240.9} & 6.07 & 241.0 & \texttt{MaxCut\_300} & \textbf{19267.0} & 0.0 & 19267.0 & \textbf{19267.0} & 0.0 & 19267.0\\
			\bottomrule[1.25pt]
		\end{tabular}
	}
	\label{tab:results}
\end{table*}

As a general conclusion, we can see through the results depicted in Table \ref{tab:results} that \texttt{QHS} clearly outperforms \texttt{LeapBQM}. While for the MCP both methods perform equally well, finding the optimum in all runs and for the entire benchmark, this trend is not replicable for the rest of the problems. For TSP, BPP, and VRP, \texttt{LeapBQM} struggles to scale as the size of the problem increases. This is not the case for \texttt{QHS}, whose scalability limits are higher than the ones studied in this paper for the TSP and BPP and whose performance is acceptable for the whole VRP benchmark. In fact, \texttt{QHS} has found the optimum value in all the executions for all the TSP, BPP, and MCP instances, as well as for these VRP cases comprised of less than seven nodes.

\section{Conclusions \& Further Work} \label{sec:conc}

The research presented in this paper provides a twofold contribution: first, we have described and categorized some of the most frequently used quantum-classical hybrid solvers. For properly doing this exercise, we have embraced the classifications and taxonomies defined in \cite{villar2023hybrid} and \cite{phillipson2023classification}. Also, we have provided a more detailed categorization according to each solver's module's contribution to both the exploration and exploitation phases. 

Secondly, we have conducted an experimentation focused on two of the solvers described: \texttt{LeapBQM} and \texttt{QHS}. For these tests, the 40 instances that comprise the QOPTLib benchmark have been employed. The results of these tests yield the general conclusion that \texttt{QHS} is ostensibly better than \texttt{LeapBQM}, showing significant superiority in a remarkable number of the evaluated instances. 

However, beyond the results, there is a much deeper reflection to be made. As previously specified, the main difference between \texttt{QHS} and \texttt{LeapBQM} lies not only in their performance but also in their composition. More specifically, both techniques differ in how they make use of quantum resources. While most hybrid solvers use quantum mechanisms for \textbf{exploitation} purposes, the \texttt{QHS} resorts to the quantum paradigm for \textbf{exploration} duties.

This change in algorithmic design may be one of the keys to the preeminence of \texttt{QHS} versus \texttt{LeapBQM}, and should make the reader reflect on the wide variety of possibilities when approaching the development of a hybrid algorithm. Indeed, algorithmic design is a crucial task not only for methods based on quantum computing \cite{koch2020demonstrating}, but also for classical techniques \cite{preitl1996algorithmic}. Thus, we encourage researchers involved in this field to devise novel hybridization mechanisms that maximize the synergies between the two computational paradigms. Undoubtedly, \textbf{the optimal placement of the puzzle pieces will unlock the secret to success in the NISQ era, and beyond}.

\section*{Acknowledgments}
This work was supported by the Basque Government through HAZITEK program (\textit{Q4\_Real}, ZE-2022/00033), through the ELKARTEK program (project \textit{KUBIT - Kuantikaren Berrikuntzarako Ikasketa Teknologikoa}, KK-2024/00105) and through Plan complementario comunicación cuántica (EXP. 2022/01341) (A/20220551). It was also supported by the Spanish CDTI through Misiones Ciencia e Innovación 2021 program (\textit{CUCO}, Grant MIG-20211005).

\bibliographystyle{splncs04}
\bibliography{biblio.bib}
\end{document}